\documentclass[apj]{emulateapj}
\usepackage{times}
\usepackage[normalem]{ulem}
\usepackage{epsfig}
\usepackage{natbib}
\usepackage{amsfonts}
\usepackage{amsmath}
\usepackage{multirow}
\usepackage{enumerate}
\usepackage{verbatim}
\usepackage[usenames]{color}
\usepackage[plainpages=false, colorlinks=true, anchorcolor=blue, linkcolor=blue, citecolor=blue, bookmarks=false]{hyperref}
\newcommand{\be}{\begin{eqnarray}}
\newcommand{\ee}{\end{eqnarray}}
\newcommand{\lp}{\left(}
\newcommand{\rp}{\right)}

\newcommand{\slugcom}{Submitted for publication in The Astrophysical Journal Letters}
\slugcomment{\slugcom}

\begin{document}

\normalsize


\title{What if the fast radio bursts 110220 and 140514 are from the same source?}

\author{Anthony L. Piro\altaffilmark{1}}
\author{Sarah Burke-Spolaor\altaffilmark{2,3}}

\altaffiltext{1}{Carnegie Observatories, 813 Santa Barbara Street, Pasadena, CA 91101, USA; piro@carnegiescience.edu}
\altaffiltext{2}{Department of Physics and Astronomy, West Virginia University, Morgantown,WV 26506, USA}
\altaffiltext{3}{Center for Gravitational Waves and Cosmology, West Virginia University, Chestnut Ridge Research Building, Morgantown, WV 26505, USA}

\begin{abstract}
The fast radio bursts (FRBs) 110220 and 140514 were detected at telescope pointing locations within 9 arcmin of each other over three years apart, both within the same 14.4 arcmin beam of the Parkes radio telescope. Nevertheless, they generally have not been considered to be from the same source because of a vastly different dispersion measure (DM) for the two bursts by over $380\,{\rm pc\,cm^{-3}}$. Here we consider the hypothesis that these two FRBs are from the same neutron star embedded within a supernova remnant (SNR) that provides an evolving DM as the ejecta expands and becomes more diffuse. Using such a model and the observed DM change, it can be argued that the corresponding SN must have occurred within $\approx10.2$ years of FRB 110220. Furthermore, constraints can be placed on the SN ejecta mass and explosion energy, which appear to require a stripped envelope (Type Ib/c) SN and/or a very energetic explosion. A third FRB from this location would be even more constraining, allowing the component of the DM due to the SNR to be separated from the unchanging DM components due to the host galaxy and intergalactic medium. In the future, if more FRBs are found to repeat, the sort of arguments presented here can be used to test the young neutron star progenitor hypothesis for FRBs.
\end{abstract}


\section{Introduction}

Fast radio bursts (FRBs) are a recently discovered class of transients characterized by millisecond bursts of radio radiation \citep{Lorimer07,Keane12,Thornton13,Ravi15}. Due to their uncharacteristically high dispersion measures (DMs), they likely occur at cosmological distances and/or in extreme density environments \citep[see discussions in][and references therein]{Kulkarni14,Luan14,Lyubarsky14,Katz16}. Furthermore, they appear to be very common, with an inferred rate of $\sim10^4$ FRBs on the sky per day \citep{Rane16}. Nevertheless, there has been no astrophysical  object or progenitor event definitively connected to FRBs, which has inspired a large number of theoretical studies to solve the mystery of identifying their progenitor. This includes magnetized neutron stars (NSs) collapsing to black holes \citep{Falcke14}, asteroids and comets falling onto NSs \citep{Geng15,Dai16}, radio flares related to soft gamma-ray repeaters \citep{Lyutikov02,Popov10,Kulkarni14,Kulkarni15,Lyubarsky14,Katz16b}, giants pulses from young pulsars \citep{Cordes16,Connor16,Lyutikov16,Popov16}, circumnuclear magnetars \citep{Pen15}, flaring stars \citep{Loeb14}, merging charged black holes \citep{Zhang16}, white dwarf mergers \citep{Kashiyama13}, and magnetic NS mergers \citep{Hansen01,Piro12,Wang16}.

Some of the strongest constraints on FRB models come from the repeating FRB 121102 \citep{Spitler14,Spitler16,Scholz16} because it is  difficult to reconcile any of the cataclysmic scenarios mentioned above with continued repetition for $>4\,{\rm yr}$. This has focused many in the community on the two scenarios of either soft gamma-ray repeater-related progenitors or giant pulse analogs from young pulsars. Furthermore, the location of FRB 121102 has now been measured \citep{Chatterjee17,Marcote17}. This allowed the host to be identified, which surprisingly was a galaxy with a low stellar mass of \mbox{$\sim(4-7)\times10^7\,M_\odot$,} reminiscent of the hosts of superluminous supernovae and long gamma-ray bursts \citep{Tendulkar17}. This is interesting because just like FRBs, magnetar-related scenarios have also often been invoked for these other types of extreme transients \citep{Metzger17}. A persistent radio source has also been identified at the location of the FRB, which could be related to the forward shock of an energetic SN or an extreme pulsar wind nebula \citep{Metzger17,Kashiyama17}.

This all begs the question though: how much of what we learn from FRB 121102 can we apply to other FRBs? Is FRB 121102 the only FRB that repeats, or do they all repeat? And if they do, why have we not found them yet? This has of course inspired efforts to follow-up the locations of known FRBs to investigate whether they can be seen to repeat \citep[e.g.,][]{Petroff15b}. Thus far there has not definitely been another repeating FRB, but there was one event that was especially interesting. FRB 140514 \citep{Petroff15a} had a DM of $562.7\,{\rm pc\,cm^{-3}}$ and was found to be within 9 arcmin of a previous FRB 110220 \citep{Thornton13} with a DM of $944.4\,{\rm pc\,cm^{-3}}$. Given the size of the Parkes error region ($\gtrsim14\,$arcmin FWHM) these two FRBs can be considered potentially colocated. \citet{Petroff15a} estimated the probability of a chance FRB at $32\%$, and so based on this non-negligible probability and the large difference in DM they argued that FRB 140514 was probably a separate source. In contrast, \citet{Maoz15} re-analyzed the probability of a chance FRB at this location based on the survey sky coverage and FRB rate. They found a probability of  $\sim1\%$ and concluded that they likely had the same source. This work then argued that the different DMs between the two bursts ruled out a cosmological intergalactic-medium origin for the DM, but might be explained by a flare-star scenario with a varying plasma blanket between bursts \citep{Loeb14}. But could this DM difference be explained by the evolution of dispersing material close to the source with the source at extragalactic distances?

Motivated by these interesting puzzles, here we study whether FRBs 110220 and 140514 could be from the same source and whether the changing DM could be due to the expansion of a SN remnant (SNR). Such a picture is naturally expected for the young NS scenario \citep{Connor16,Lyutikov16,Piro16} as well as some magnetar scenarios if the magnetar is young  \citep{Metzger17}. In Section \ref{sec:remnant}, we present a toy model for the SNR evolution and describe what constraints can be placed by comparison to FRBs 110220 and 140514. In Section \ref{sec:repeater}, we apply many of the same arguments to the repeating FRB 121102. In \mbox{Section \ref{sec:discussion},} we summarize our main results and discuss potential implications that stem from our work.

\section{Constraints from FRBs 110220 and 140514}
\label{sec:remnant}

\subsection{Supernova Remnant Model}

As an SNR expands and cools, the material initially recombines over the timescale of $\sim\,$months to about a year. Soon after though, the interaction of the SNR with the interstellar medium (ISM) creates a reverse shock that passes back through the ejecta. This shock reaches temperatures sufficient to ionize the material, producing free electrons that can now once again disperse radio emission \citep{Piro16}. 

Assuming that the Milky Way component of an FRB's DM can be subtracted out, the remaining total observed DM is
\be
	{\rm DM}_{\rm tot}(t) = \frac{{\rm DM}_{\rm SNR}(t)}{1+z} + \frac{{\rm DM}_{\rm host}}{1+z} + {\rm DM}_{\rm IGM},
\ee
where ${\rm DM}_{\rm SNR}(t)$, ${\rm DM}_{\rm Host}$, and ${\rm DM}_{\rm IGM}$ are the components to the DM from the SNR, host galaxy, and intergalactic medium (IGM), respectively, and $z$ is the redshift to the source. The IGM component is given by
\be
	{\rm DM}_{\rm IGM} = \frac{n_0c}{H_0}z,
\ee
where $H_0$ is Hubble's constant and $n_0=1.6\times10^{-7}\,{\rm cm^{-3}}$ is the present-day density assuming the baryons are homogeneously distributed and ionized \citep{Katz16c}. At its most simplistic level, an SNR with mass $M$ and expanding with a velocity $v$ provides a time dependent DM of
\be
	{\rm DM}_{\rm SNR}(t) &\approx& \frac{3fM}{4\pi (vt)^2\mu_e m_p}
	\nonumber
	\\
	&=& 9.4\times10^4\mu_e^{-1}v_9^{-2}t_{\rm yr}^{-2}\lp\frac{fM}{M_\odot} \rp{\rm pc\,cm^{-3}},
	\label{eq:DM snr}
\ee
where $f$ is the fraction of the ejecta that has been ionized by the reverse shock \citep[note that an energetic central source, such as a magnetar, could contribute to ionizing the ejecta from the inside-out,][]{Metzger17}, $m_p$ is the proton mass, $\mu_e$ is the mean molecular weight per electron, $v_9=v/10^9\,{\rm cm\,s^{-1}}$, and $t_{\rm yr}=t/1\,{\rm yr}$. For a pure hydrogen composition, $\mu_e=1$, for a solar composition $\mu_e\approx1.2$, and for a heavier composition (as in a stripped-envelope SN) $\mu_e\approx 2$.

  \begin{deluxetable}{lcc}
  \tablecolumns{10} \tablewidth{200pt}
 \tablecaption{Comparison of the FRBs 110220 and 140514}
   \tablehead{ Name & FRB 110220\tablenotemark{a} & FRB 140514\tablenotemark{b}}
  \startdata
   Event date UTC & 20 Feb 2011 & 14 May 2014 \\
   Event time UTC & 01:55:48.96  & 17:14:11.06 \\
   RA & 22$^{\rm h}$34$^{\rm m}$38$^{\rm s}$\ & 22$^{\rm h}$34$^{\rm m}$06$^{\rm s}$\\
   Dec & -12$^{\circ}$24' & -12$^{\circ}$18' \\
   ${\rm DM}_{\rm FRB}$ (${\rm pc\,cm^{-3}}$) & 944.4 & 562.7 \\
   ${\rm DM}_{\rm MW}$ (${\rm pc\,cm^{-3}}$) &  34.9 & 34.9 \\
   Observed width (ms) & 5.6 & 2.8 \\
   Peak flux density (Jy) & 1.3 & 0.47 \\
   Fluence (Jy ms) & 8.0 & 1.3
   \enddata
      \tablenotetext{a}{\citet{Thornton13}}
        \tablenotetext{b}{\citet{Petroff15a}}
\label{table}
\end{deluxetable}

The unchanging contributions to the DM from the host and IGM make it difficult to tease out the SNR contribution. One way to overcome this is by considering the change in DM instead, since these constant factors will stay the same. Calculating the integral from times $t_1$ to $t_2$, one finds
\be
	\Delta {\rm DM}_{\rm tot} =  -\frac{3fM}{4\pi v^2\mu_e m_p} \lp \frac{1}{t_1^2} - \frac{1}{t_2^2}\rp (1+z)^{-1}.
\ee
Taking $t_1=t$ and $t_2=t+\Delta t$, in the limit that $t\gg\Delta t$, then
\be
	\Delta {\rm DM}_{\rm tot} &\approx& -\frac{3fM}{2\pi (vt)^2\mu_e m_p}\frac{\Delta t}{t}(1+z)^{-1}
	\nonumber
	\\
	&=& -\frac{1.9\times10^5}{1+z}\mu_e^{-1}v_9^{-2}t_{\rm yr}^{-2}\frac{\Delta t}{t}\lp\frac{fM}{M_\odot} \rp {\rm pc\,cm^{-3}}.
	\nonumber 
	\\
	\label{eq:delta DM}
\ee
can be used as an approximation for the change in DM.

For FRBs 110220 and 140514, the decrease in DM was $381.7\,{\rm pc\,cm^{-3}}$ on a timescale of $\Delta t=3.2\,{\rm yrs}$, as can be seen from Table \ref{table} along with other properties of these FRBs. Using Equation (\ref{eq:delta DM}) we can then estimate roughly when the SN associated with FRBs 110220 and 140514 occurred as a function of the mass of the ejecta, resulting in
\be
	t\approx 11.6(1+z)^{-1/3}\mu_e^{-2/3}v_9^{-2/3} \lp\frac{fM}{M_\odot} \rp^{1/3}{\rm yr}.
\ee
Thus the SN must have occurred rather recently to explain the observed change in DM.

\subsection{Constraints on Explosion Time, Ejecta Mass, and Supernova Properties}

In fact, even more stringent, model-independent constraints on the explosion time can be made by considering the ratio of the DMs. This is given by
\be
	\frac{\displaystyle\frac{{\rm DM}_{\rm SNR}}{(1+z)(t_{\rm yr}+3.2)^2}+\frac{{\rm DM}_{\rm host}}{1+z} + {\rm DM}_{\rm IGM}}{\displaystyle\frac{{\rm DM}_{\rm SNR}}{(1+z)t_{\rm yr}^2}+\frac{{\rm DM}_{\rm host}}{1+z} + {\rm DM}_{\rm IGM}}
	= \frac{527.8}{909.5} = 0.58,
	\nonumber
	\\
\ee
where we have subtracted out the Milky Way component from each DM and on the left-hand side we assume that ${\rm DM}_{\rm SNR}$ is evaluated at $1\,{\rm yr}$ after the SN. The thing to note is that the non-zero contributions of ${\rm DM}_{\rm host}$ and ${\rm DM}_{\rm IGM}$ always push the ratio closer to unity. Thus independent of the details of the SN, this ratio must obey the limit that
\be
	\frac{t_{\rm yr}^2}{(t_{\rm yr}+3.2)^2} < 0.58\hspace{0.25cm} \Rightarrow  \hspace{0.25cm} t<10.2\,{\rm yrs}.
	\label{eq:timelimit}
\ee
Thus if these FRBs are associated with a young NS, the SN must have taken place less than $10.2\,{\rm yrs}$ before FRB 110220.

Given this more model-independent constraint on the explosion time, we can turn this around and put a constraint on the SN ejecta mass. Basically, if the SN happened so recently, the ejecta cannot be too massive. Otherwise, it would provide too large of a DM. Taking Equation (\ref{eq:DM snr}), setting ${\rm DM}_{\rm SNR}=909.5(1+z){\rm pc\,cm^{-3}}$, and using the time limit given by Equation (\ref{eq:timelimit}) results in
\be
	M < 1.0(1+z) f^{-1} \mu_e v_9^2\, M_\odot. 
	\label{eq:mass_v}
\ee
This can also be related to the energy of the explosion using $E\approx Mv^2/2$ resulting in
\be
	M < 1.0(1+z)f^{-1/2}\mu_e^{1/2} E_{51}^{1/2}\,M_\odot,
	\label{eq:mass}
\ee
where $E_{51}=E/10^{51}\,{\rm erg}$.

\begin{figure}
\epsscale{1.2}
\plotone{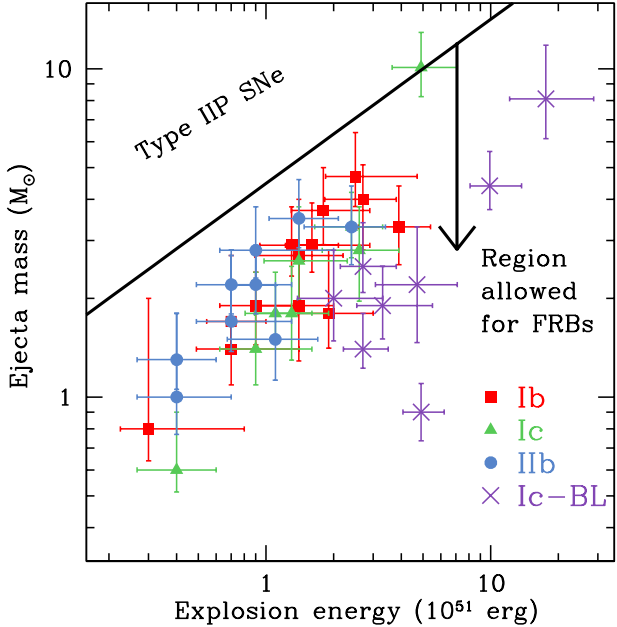}
\caption{Stripped envelope SN ejecta masses and explosion energies estimated by \citet{Lyman16} for 38 different events. The colors and symbols denote Type Ib (events that show helium; red squares), Type Ic (events that show no helium; green triangles), Type IIb (events that show helium and trace hydrogen at early times; blue circles), and Type Ic-BL (events that show no helium and are especially energetic; purple cross). The solid black line is the limit placed by Equation (\ref{eq:mass}) using $f=0.1$, $\mu_e=2$, and $z\ll1$. The upper left corner marked ``Type IIP SNe'' labels the parameter regime roughly found for hydrogen-rich SNe by \citet{Pejcha15}. Note that when comparing to these events that $\mu_e\approx1.2$ and thus the black lines should be moved down by a factor of $\approx1.3$.}
\label{fig:sn_ejecta}
\epsscale{1.0}
\end{figure}

We write the mass constraint in this way because it helps in comparing to SN observations. In particular, \citet{Lyman16} use the light curves of 38 stripped envelope SNe (explosions that have had all or most of their hydrogen stripped) to estimate the ejecta masses and explosion energies. This is summarized in Figure \ref{fig:sn_ejecta} with the various colors and symbols designating the various subclasses of events. In comparison, we show the constraint placed by Equation (\ref{eq:mass}), where we take $f=0.1$ (only a small amount of ionization given the early stages inferred by the time limit) and $\mu_e=2$ (appropriate for SNe of these types). This comparison shows that these stripped-envelope SNe naturally have the ejecta masses and energetics consistent with this limit.

In contrast, we also label the region that roughly corresponds to hydrogen-rich SNe as ``Type IIP SNe'' in the upper left corner \citep[using the work of][]{Pejcha15}. This shows that a hydrogen-rich SN just has too much mass around to explain the DM for these FRBs \citep[as found in the more detailed discussion by][]{Piro16}. Especially note that hydrogen-rich material has more electrons per unit mass than the hydrogen deficient material of a stripped SN, thus for $\mu_e\approx1.2$ the solid line in Figure \ref{fig:sn_ejecta} should move down by a factor of $\approx1.3$.

A natural question to ask is whether a connection can be made between the SN constraints made here and the unique galaxy host found for the repeating FRB 121102. For a stellar mass of $\sim(4-7)\times10^7\,M_\odot$  \citep{Tendulkar17} and the normal mass-metallicity relation \citep{Tremonti04}, one would expect the host to have a very sub-solar metallicity (although the currently available limits only constrain it to be solar metallicity or below). In contrast, Type Ib and Ic SNe happen preferentially in hosts with larger metallicities than Type II \citep[e.g.,][]{Prieto08}. The Type Ic-BL that do not have associated gamma-ray bursts also follow regions with a similarly high metallicity \citep[e.g.,][]{Modjaz08,Modjaz11}. The main events that have a preference for low metallicity environments are hydrogen-deficient superluminous SNe and long gamma-ray burst \citep[e.g.,][]{Perley16a,Perley16}, as has been pointed out by \citet{Metzger17}. Such events would have parameters that roughly follow the Type Ic-BL points in Figure \ref{fig:sn_ejecta}, and thus be consistent with the limits we find here.

Another possibility is that the connection between the host galaxy and the FRB is not the presence of a magnetar but rather a quickly spinning progenitor. The low metallicity environments are commonly thought to inhibit mass and angular momentum loss, allowing a massive star to spin much more quickly near the end of its life as needed for generating engine-driven explosions \citep[e.g.,][]{Yoon05}. In fact, \citet{Piro16} argues that it is preferential to have NSs born with a fast spin ($\lesssim3\,{\rm ms}$), but a relatively normal magnetic field ($\lesssim10^{12}\,{\rm G}$)  if FRBs are powered by the NS rotation. This is because a high magnetic field would cause the NS to spin down too quickly when the SN remnant is still too optically thick. In this picture, long gamma-ray bursts and superluminous SNe may represent the most extreme cases where a dynamo is able to generate an especially large field, while FRBs are from progenitors with lower fields. Nevertheless, in each case the connection to the low metallicity host is a high spin. This might also help explain the high rate of FRBs, since both long gamma-ray bursts and superluminous SNe occur at rates too small by orders of magnitude in comparison to FRBs.

\subsection{Absorption, Dispersion, and Scattering Constraints}
\label{sec:constraints}

There are other important constraints to consider for an SNR to be able to explain the changing DM between FRBs 110220 and 140514. The first is that the radio emission should not be inhibited by free-free absorption \citep{Piro16}. For radiation at frequencies $h\nu\ll k_{\rm B}T$, where $T$ is the temperature, the absorption coefficient is \citep{Rybicki79}
\be
	\alpha_{\rm ff} = 1.9\times10^{-2}T^{-3/2} Z^2n_en_i\nu^{-2}g_{\rm ff}\,{\rm cm^{-1}},
\ee
where $Z$ is the average charge per ion, $n_e$ and $n_i$ are the electron and ion number densities, respectively, $g_{\rm ff}\sim1$ is the Gaunt factor, and all quantities are in cgs units. Estimating $n_e=3fM/3\pi(vt)^3\mu_e m_p$ and $n_i=3fM/3\pi(vt)^3\mu_i m_p$, where $\mu_i$ is the mean molecular weight be ion, and assuming the characteristic size to be $vt$  results in a limit on the mass for $t<10.2\,{\rm yrs}$ of
\be
	M<2.0(1+z)T_4^{3/4}v_9^{5/2}\frac{(\mu_e\mu_i)^{1/2}}{fZ}\nu_{1.4}\,M_\odot,
\ee
where $T_4=T/10^4\,{\rm K}$ and $\nu_{1.4}=\nu/1.4\,{\rm GHz}$. Note the redshift factor of $1+z$ is needed since $\nu$ is the observed frequency and what matters for the absorption is the frequency in the frame of the source. Again this constraint appears to argue for a lower mass like a stripped-envelope SN rather than a massive hydrogen-rich SN. This estimate is what is needed for the FRB to be observed at $1.4\,{\rm GHz}$. At lower frequencies, the FRB may still be absorbed. For example, if the mass marginally obeys the limit placed by Equation (\ref{eq:mass_v}), the one would expect the FRB to be absorbed for frequencies
\be
	\nu < 700 (1+z)^{-1} ZT_4^{-3/4}v_9^{-1/2}\lp\frac{\mu_e}{\mu_i} \rp^{1/2}{\rm MHz}.
\ee
Interestingly, even though there have been searches for FRBs by the Low-Frequency Array, the Murchison Widefield Array, and the Green Bank Telescope at low frequencies, an FRB has never been seen below $700\,{\rm MHz}$. Could this be explained by the presence of an SNR?

Another constraint is that the observed dispersion for these FRBs is very close to $\nu\propto t^{-\alpha}$ with $\alpha\approx2$. At high densities, the $\alpha$ can be slightly higher than $2$ \citep{Katz16}
\be
	\alpha =  2 + \frac{3e^2n_e}{2\pi m_e\nu^2}.
\ee
For the mass limit we find, this additional factor is $\sim10^{-6}$ and thus we do not expect the SNR to have a strong impact on the time dependence of the dispersion.

The typical pulse width of FRBs is $\sim5\,{\rm ms}$, too broad to be explained by Kolmogorov turbulence in the IGM \citep{Luan14}. Furthermore, galactic pulsars show that  scattering in the interstellar medium cannot contribute significantly to the pulse widths of FRBs \citep{Krishnakumar15}. The dense environment of an SNR might provide the necessary scattering. The delay attributed to scattering by an angle $\Delta\theta$ is \citep{Williamson72,Kulkarni14,Katz16c}
\be
	W \approx \frac{vt}{2c}(\Delta\theta)^2.
\ee
For a timescale of $t\sim10\,{\rm yr}$, this would require a scattering of $\Delta \theta\sim10^{-5}$. As more FRBs are found and constraints made on the time since an associated SN, a better census on the scattering angles can be made.

\subsection{Where is the Supernova?}

\begin{figure}
\epsscale{1.2}
\plotone{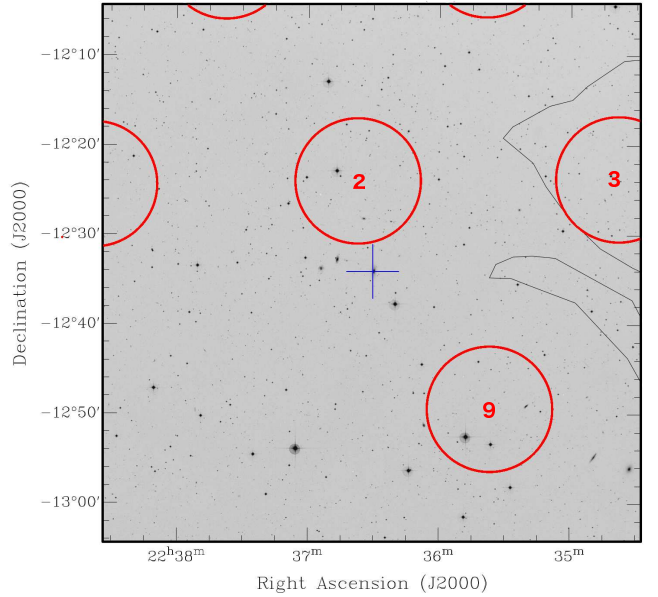}
  \caption{The position of SN 2001hh (shown with a blue cross) in comparison to the Parkes beams at the time of the detection of the brighter FRB 110220 (beam positions and names shown with red circles). The bright beam 3 detection, coupled with the non-detection of the burst in adjacent beams 1, 2, 4, 9, and 10, places limits on the position of the burst (50\% confidence limit shown as black bounds). If the FRB source was SN 2001hh, we would have expected to detect the FRB most significantly in beam 2 rather than beam 3.}
  \label{fig:sn2001hh}
\epsscale{1.0}
\end{figure}

The constraints that can be placed on the explosion time of a SN related to FRBs 110220 and 140514 naturally beg the question, where is the SN? In fact, looking through the SN archives we found that SN 2001hh took place in 2001 \citep{Hakobyan08}, roughly on the same part of the sky as FRBs 110220 and 140514 and with a host galaxy redshift of $z\sim0.02$, which is consistent with the DM-inferred upper redshift limit of FRB 140514 of $z\lesssim0.5$ \citep{Petroff15a}. A detailed comparison of the locations is shown in Figure \ref{fig:sn2001hh}, which demonstrates that in fact SN 2001hh is roughly 38 arcmin away from the pointing location of FRB 110220. The location of the supernova during the FRB 110220 detection indicates that if this supernova and the FRB events were related, that FRB 110220 would have been detected in beam 2 rather than beam 3, as it was. Thus, the location and timing of SN 2001hh is a funny coincidence, but does not appear associated with these FRBs. Furthermore, SN 2001hh was a Type II and may have had too much mass associated with it to explain the observed DM anyway.


The fact that a SN was not found associated with FRBs 110220 and 140514 does not rule out this model. In 2001, our ability to find all nearby SNe was much less than it is now, and it is quite reasonable that this SN was not found. This example does show that how going into the future, we should be able to compare the locations of FRBs with nearby SNe to test whether the young NS scenario makes sense for these events.

\section{The Repeating FRB 121102}
\label{sec:repeater}

Similar arguments as presented above can also be applied to the repeating FRB 121102. In this case, the DM is not seen to change with time, so we can only place limits. Using the fact that this FRB has been observed for $\approx4\,{\rm yrs}$ and the DM is observed to be the same within $\approx3\,{\rm pc\,cm^{-3}}$, we can use Equation (\ref{eq:delta DM}) to constrain the explosion time to be
\be
	t> 60 v_9^{-2/3} (1+z)^{-1/3} \lp\frac{fM}{M_\odot} \rp^{1/3} \lp \frac{\Delta t}{4\,{\rm yr}} \rp^{1/3}{\rm yr}.
\ee
Furthermore, since this FRB's distance is known, subtracting off the Milky Way and IGM contributions to the DM gives the rough limit that ${\rm DM}_{\rm SNR}\lesssim 225(1+z){\rm pc\,cm^{-3}}$ \citep{Tendulkar17}. Using Equation (\ref{eq:DM snr}), the ejecta mass must then satisfy
\be
	M\lesssim 8 (1+z)f^{-1}v_9^2 \lp \frac{t}{60\,{\rm yr}}\rp^2 M_\odot.
\ee
Thus we find that the repeating FRB 121102 is not inconsistent with a young NS origin as long as the SN occurred sufficiently long ago. The mass constraints we find are not too dissimilar than a SN, but not as constraining as what was found for FRBs 110220 and 140514. As we continue to observed bursts from FRB 121102 and measure the associated DMs, we will be able to place tighter constraints on any potential SNR around this FRB.

\section{Discussion and Conclusions}
\label{sec:discussion}

We have considered the hypothesis that FRBs 110220 and 140514 are from the same source with the difference in DM between the two busts (of greater than $380\,{\rm pc\,cm^{-3}}$) due to the expansion of an SNR. From this, we can place constraints that the corresponding SN must have occurred $<10.2\,{\rm yr}$ before FRBs 110220 and that the ejecta mass of the SN must be relatively low or the SN be very energetic, similar to what is measured for stripped-envelope SNe. We in fact found a SN at a similar time and location on the sky (SN~2001hh), but its distance of $\sim$\,38 arcmin from the nominal location of FRB 110220 means that the SN is probably not related. The observation of a third FRB from this location would be even more constraining, since this would allow the IGM and host components of the DM to be separated from the SNR component. With a reasonable estimate for the host component, the DM from the IGM would allow the distance to these FRBs to be estimated to hopefully narrow down which galaxies could potentially be the host of these FRBs.

Whether or not it ends up being true that FRBs 110220 and 140514 are from the same source, this work highlights the various constraints that can be placed in the future as hopefully more FRBs are seen to repeat. Even in the case of FRB 121102, which has been seen to repeat but a change in the DM have not been observed, we have shown that interesting constraints can be placed on the young NS hypothesis for FRBs.

Going into the future, there are a few key implications for future observations that stem from this work. 

First, the potential for a time-dependent change in dispersion may impact searches for repeating FRBs. Searches that rely on multiple events appearing at one DM (say, to look for excesses at a given DM over observations at month to year spacings, or follow-up searches that rely on coherent dedispersion to raise sensitivity to a fixed DM) will not uncover repeating events associated with the youngest SNe, for which rapidly-changing dispersion measures can occur.

Second, finding a low frequency cutoff due to free-free absorption, as discussed in Section \ref{sec:constraints}, would be important for probing the environment around FRBs. This cutoff should move to lower frequencies as the DM decreases. In the future, the  Canadian Hydrogen Intensity Mapping Experiment \citep[CHIME,][]{chime} will be ideally suited to do this since it will collect a large number of FRBs and be sensitive to a frequency range of $400-800\,{\rm MHz}$ where this free-free absorption is expected to occur. This result further highlights the fact that very low frequency observations ($\lesssim$500\,MHz) will only probe a much older, and potentially less energetic, population of spinning compact objects.

Finally, as more FRBs are found, their locations should be compared with nearby SNe to test whether the young NS scenario makes sense for these events. This will be  aided by surveys with rapid cadences that will increase our efficiency for discovering SNe, such as the All-sky Automated Survey for Supernovae \citep[ASAS-SN,][]{shappee:14} and the Zwicky Transient Facility \citep[ZTF,][]{law:09}, as well as the Large Synoptic Survey Telescope \citep{lsst}, which will provide a comprehensive record of almost every SN at larger distances. FRBs with especially high DMs might be important in this sense if their large DMs indicate they are being found soon after the explosion.

\acknowledgments
We thank Edo Berger, Liam Connor, and Emily Petroff for helpful conversations, and Benjamin Shappee for assistance in finding SN 2001hh. S.B.S. has been supported by NSF award \#1458952. We acknowledge partial support from the Research Corporation for Scientific Advancement (RCSA) for participation in the meeting Fast Radio Bursts: New Probes of Fundamental Physics and Cosmology at the Aspen Center for Physics (February 12-17, 2017) where much of this work was inspired.

\bibliographystyle{apj}

\end{document}